\documentstyle[12pt,pra,aps,psfig]{revtex}
\tightenlines
\begin{document}
\draft

\title{\LARGE \bf Nonclassical correlations\\
  in damped $N$-solitons}

\author{Eduard~Schmidt,  Ludwig~Kn\"{o}ll, and Dirk--Gunnar~Welsch}
\address{Friedrich-Schiller-Universit\"{a}t Jena,
Theoretisch-Physikalisches Institut\\
Max-Wien Platz 1, D-07743 Jena, Germany}

\maketitle

\begin{abstract}
The quantum statistics of damped higher-order optical solitons
are analyzed numerically, using cumulant-expansion techniques in
Gaussian approximation. A detailed analysis of nonclassical properties
in both the time and the frequency domain is given,
with special emphasis on the role of absorption.
Highly nonclassical broadband spectral correlation
is predicted.
\end{abstract}

\vspace{8mm}

\thispagestyle{empty}

  From classical optics it is well known that
nonlinearities
can compensate for the dispersion-assisted pulse spreading
\cite{HasegawaA89,AkhmanovSA92} or for diffraction-assisted
beam broadening (see, e.g., \cite{AitchisonJS91}). In the
two cases, the
undamped
motion of the (slowly varying) bosonic field variables
$\hat{a}(x,t)$ is governed
by the Hamiltonian
\vspace*{-.5ex}
\begin{eqnarray}
\label{eq.aa}
&
\hat{H}=\hbar \int dx\left[
{\textstyle\frac{1}{2}}\omega ^{(2)}\big( \partial
_{x}\hat{a}^{\dagger }\big) \big( \partial _{x}\hat{a}\big)
+{\textstyle\frac{1}{2}}\chi
\hat{a}^{\dagger }\hat{a}^{\dagger }\hat{a}\hat{a}
\right] ,
&\\[.5ex]&
\left[ \hat{a}(x,t),\hat{a}^{\dagger }(x^{\prime },t)\right] =
\delta (x-x^{\prime })
&
\label{eq.aa1}
\end{eqnarray}

\vspace{-.5ex}
\noindent
[$t$, propagation variable; $x$, ``transverse'' coordinate;
$\omega^{(2)}$, second order dispersion or diffraction constant;
${\chi }$ nonlinearity constant;
see, e.g., \cite{MecozziA98,SchmidtE99}].
Note that bright temporal solitons can be formed either
in focusing media with anomalous dispersion
(\mbox{$\chi$ $\!<$ $0$}, \mbox{$\omega^{(2)}$ $\!>$ $\!0$}) or
in defocusing media with normal dispersion
(\mbox{$\chi$ $\!>$ $\!0$}, \mbox{$\omega^{(2)}$ $\!<$ $\!0$}),
whereas spatial solitons require always focusing nonlinearity.
The effect of absorption is described in terms of ordinary
Markovian relaxation theory
resulting, in the low temperature limit,
in the master equation
\vspace*{-1ex}
\begin{equation}
i\hbar ~\partial _{t}\hat{\rho}=[\hat{H},\hat{\rho}]
+i\gamma
\hbar \int dx
 \left( 2\hat{a}\hat{\rho}\hat{a}^{\dagger }
  -\hat{\rho}\hat{a}^{\dagger }\hat{a}
  -\hat{a}^{\dagger}\hat{a}\hat{\rho}\right)
\label{eq.master}
\end{equation}

\vspace{-1ex}
\noindent
($\gamma$, damping constant).

The master equation (\ref{eq.master}) is converted,
after spatial discretization, into a pseudo-Fokker-Planck equation
for an $s$-parametrized multi-dimensional phase-space
function, which is solved numerically using
cumulant expansion in Gaussian approximation \cite{SchmidtE99}.
The initial condition is realized by a multimode coherent
state without internal entanglement,
and it is assumed that the field expectation value
corresponds to the classical
$N$-soliton solution,
$\langle \hat{a}(x,t_0)\rangle$ $\! =$ $\! N \,a_0\, $ $\!\mbox{sech}(x/x_0)$,
$N$ $\!=$ $\!1,2,\ldots$ ($a_0$ and $x_0$, mean amplitude and
width of the fundamental soliton, respectively).

Spectral properties can be studied introducing the Fourier-component
operators
\vspace*{-.5ex}
\begin{equation}
\hat{a}(\omega ,t)=(2\pi)^{-\frac{1}{2}}\int_{-\infty }^{\infty }dx\,
e^{i\omega x}\hat{a}(x,t).
\label{eq.aw}
\end{equation}

\vspace{-.5ex}
\noindent
Here we restrict our attention to correlations of photon number
fluctuations. In the case of fiber soliton pulses the correlations in the
$\omega$-domain can be measured using appropriate spectral filtering
(see, e.g., in \cite{SpaelterS98}).
In the case of spatial solitonic beams the correlations in both
the $x$ and $\omega$-domains, respectively, can be measured
by filtering
the field in the near- and far-field zones
of the output beam (see, e.g., \cite{MecozziA98}).

\pagebreak
\noindent
\begin{figure}[tbh]
\begin{minipage}{6.2in}
\par
 \hbox{\centerline{\psfig{file=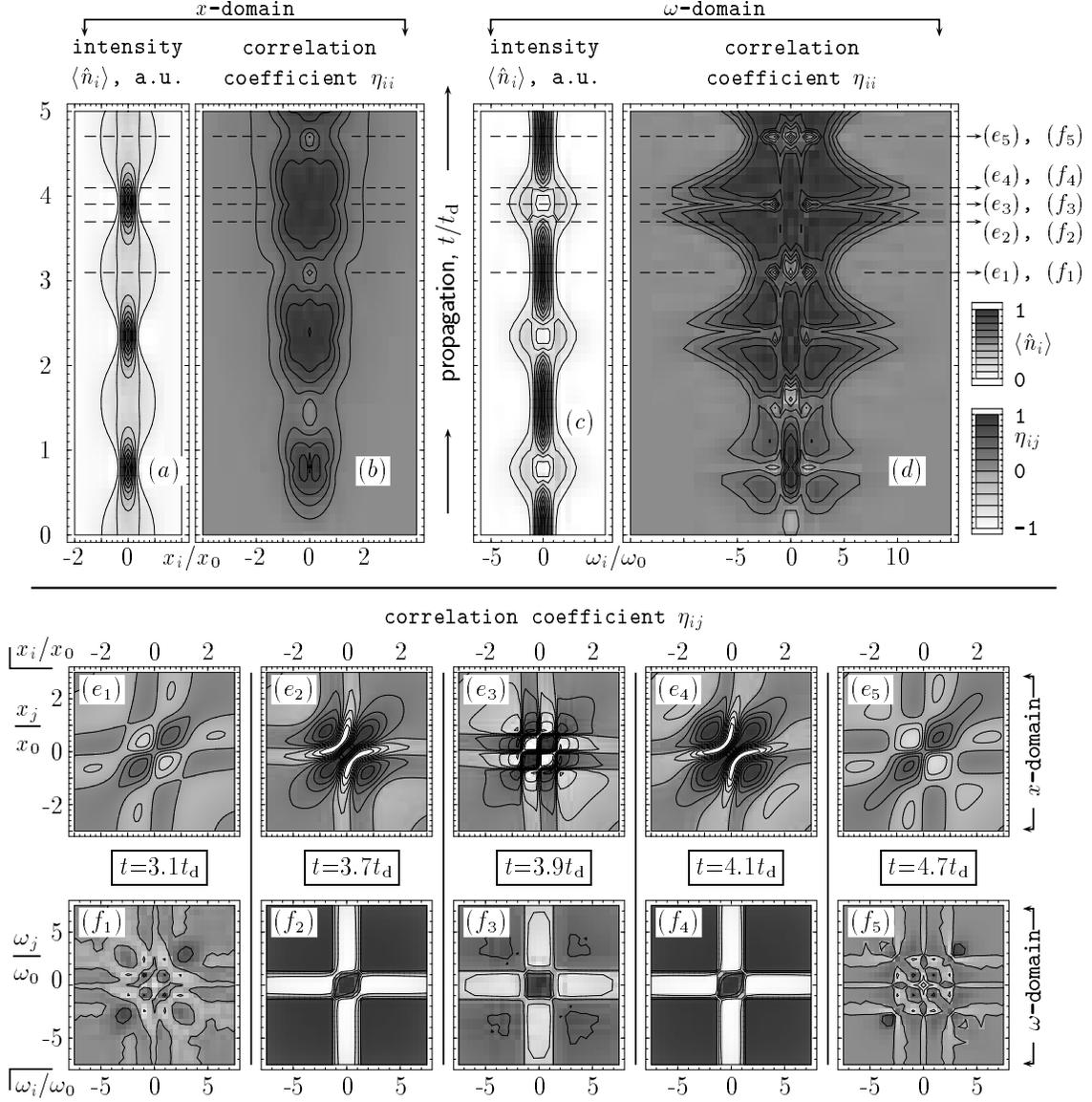,width=6.0in,bbllx=58pt,bblly=140pt,bburx=552pt,bbury=644pt}}}
\par
\caption{
The evolution of the mean photon number $\langle\hat{n}_i\rangle$
and the correlation coefficient $\eta_{ii}$
of an undamped soliton, $N$ $\!=$ $\!2$,
is plotted in the $x$-domain, $\Delta x$ $\!=$ $\!0.05\,x_0$ [$(a)$,$(b)$],
and the $\omega$-domain, $\Delta \omega$ $\!=$ $\!0.25\,\omega_0$
[$(c)$,$(d)$]. The plots $(e_1)\!-\!(e_5)$
($x$-domain) and $(f_1)\!-\!(f_5)$ ($\omega$-domain)
show the correlation coefficient $\eta_{ij}$ for
typical propagation lengths ($\omega_0$ $\!=$ $\!1/x_0$,
$t_{\rm d}\!=\!|x_0^2/\omega^{(2)}|$,
$\int dx \, \hat{a}^{\dagger }(x ,0)\hat{a}(x ,0)$ $\!=$ $\!8\times10^9$).
}
\label{fig1}
\end{minipage}
\end{figure}
%
\vspace*{2ex}

The output can be given by
(see, e.g., \cite{LevandovskyD99})
\begin{eqnarray}
  \hat{b}(\nu,t)=G(\nu,t)\hat{a}(\nu,t)
    +\sqrt{1-|G(\nu,t)|^2}\,\,\hat{f}(\nu,t),
\label{eq.filter}
\end{eqnarray}
where, according to the domain considered, $\nu$ stands for
$x$ or $\omega$, and $G(\nu,t)$,
$|G(\nu,t)|$ $\!\le$ $\!1$, is the (complex) transmittance of the
filter and $\hat{f}(\nu,t)$ is a bosonic noise operator.
The photon number operator of the detected light is
$\hat{n}$ $\!=$ $\!\int d\nu \, \hat{b}^{\dagger }(\nu ,t)\hat{b}(\nu ,t)$.
Assuming square bandpass filters with $G_i(\nu,t)$ $\!=$ $1$ if
$|\nu$ $\!-$ $\!\Omega_i|$ $\!\leq$ $\!\Delta\Omega$ and
$G_i(\nu,t)$ $\!=$ $\!0$ otherwise, we consider the correlation
coefficient
%
\begin{center}
\begin{minipage}{6.2in}
\begin{figure}[tbh]
\par
 \hbox{\centerline{\psfig{file=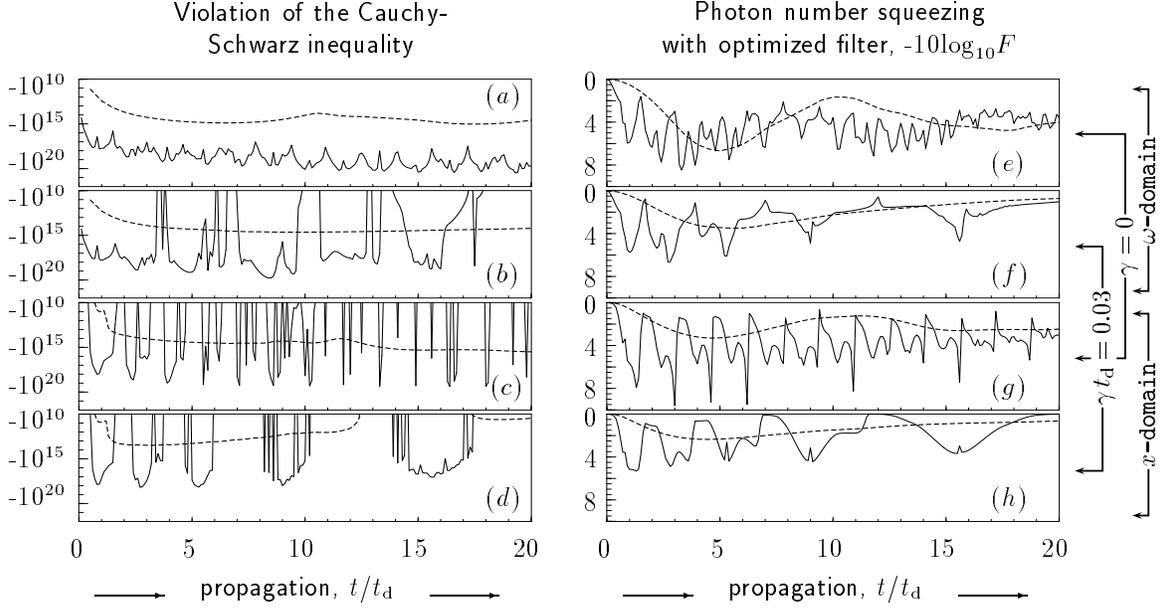,width=6.0in,bbllx=55pt,bblly=370pt,bburx=555pt,bbury=644pt}}}
\par
\caption{
The maximal violation of the Cauchy--Schwarz inequality
for the photon number fluctuation
[plots $(a)\!-\!(d)$] and
the smallest Fano factor
$F$ $\!=$ $\!\langle\Delta \hat{n}^2\rangle/\langle \hat{n}\rangle$
(strongest photon number squeezing)
achievable with optimized filters
[plots $(e)\!-\!(h)$] are shown for the fundamental soliton, $N$ $\!=$ $\!1$,
(dotted line) and the soliton with $N$ $\!=$ $\!2$ (full line)
[$x$-domain: plots $(a),(b),(e),(f)$,
$\omega$-domain: plots $(c),(d),(g),(h)$;
$\gamma$ $\!=$ $\!0$: plots $(a),(c),(e),(g)$,
$\gamma t_{\rm d}$ $\!=$ $\!0.03$: plots $(b),(d),(f),(h)$;
other parameters as in Fig.~\protect\ref{fig1}].
}
\label{fig2}
\end{figure}
\end{minipage}
\end{center}
%
\vspace*{2ex}

\begin{eqnarray}
\eta_{ij}&=&\frac
      {\langle:\!\Delta\hat{n}_i\Delta\hat{n}_j\!:\rangle}
      {\sqrt{\langle\Delta\hat{n}_i^2\rangle\langle\Delta\hat{n}_j^2\rangle}}
=\frac{c_{ij}}{\sqrt{(c_{ii}+m_i)(c_{jj}+m_j)}}
\label{eq.etaij}
\end{eqnarray}
($m_i$ $\!=$ $\!\langle \hat{n}_{i}\rangle$,
$c_{ij}$ $\!=$ $\!\langle: \!\Delta \hat{n}_{i} \Delta \hat{n}_{j}\!:\rangle$,
$\Delta \hat{n}_{i}$ $\!=$ $\!\hat{n}_i-m_i$), where
$:\ :$ introduces normal ordering.
It can be shown that $\eta _{ii}$ $\!\leq$ $\!1$, and
$|\eta _{ij}|$ $\!\leq$ $\! 1$ for nonoverlapping intervals.
A negative sign of the coefficient $\eta _{ii}$ or
a value smaller than unity of the Fano factor
$F_i$ $\!=$ $\!\langle\Delta \hat{n}_{i}^2\rangle/\langle \hat{n}_i\rangle$
$\!=$ $\!(1$ $\!-$ $\!\eta_{ii})^{-1}$
indicates photon number squeezing
of the filtered light.

   From Fig.~\ref{fig1} it is seen that typical
changes in the evolution of $\langle\hat{n}_i\rangle$
[Figs.~\ref{fig1}$(a),(c)$] and those of $\eta_{ii}$
[Figs.~\ref{fig1}$(b),(d)$] and
$\eta_{ij}$ [Figs.~\ref{fig1}$(e_1)\!-\!(e_5),(f_1)\!-\!(f_5)$] are
closely related to each other. Near the points of soliton compression
[maxima of $\langle\hat{n}_i\rangle$ in Fig.~\ref{fig1}$(a)$]
the formation of strong-correlation patterns is observed
[Figs.~\ref{fig1}$(e_2)\!-\!(e_4),(f_2),(f_4)$].
In contrast to the $x$-domain [Fig.~\ref{fig1}$(b)$],
sub-Poissonian statistics is observed in the $\omega$-domain
[Fig.~\ref{fig1}$(d)$]. Moreover, the correlation in the
$\omega$-domain extends over a larger interval
(relative to the corresponding initial pulse width)
than the correlation in the $x$-domain.
One possible explanation of
such strong, almost perfect correlation ($|\eta_{ij}|$ $\!\to$ $\!1$)
can be seen in the instability of the classical $N$-soliton solution. From
a linearization approach \cite{HausHA90}, the internal noise
of a quantum soliton should be associated with interferences
\cite{MecozziA97} between the soliton components and the continuum part
of the solution to the classical nonlinear Schr\"odinger equation, as
obtained by means of inverse scattering method (see, e.g.,
\cite{HasegawaA95}). The qualitative changes observed for
turning from the fundamental soliton to higher-order solitons
($N=1$ $\!\to$ $\! N$ $\!=$ $\!2,3,\ldots$) are due to the
presence of {\it more than one} soliton component.
Discrepancies between the parameters (amplitude, group velocity, etc.)
of the soliton components of the $N$-soliton solution
play the central role in establishing very strong
internal correlations.

Nonclassical correlation can be detected, e.g.,
by testing the Cauchy-Schwarz inequality for the normally
ordered photon number variances. When it is violated, i.e.,
\begin{equation}
c_{ii}c_{jj}-c_{ij}^2 < 0 ,
\label{eq.CS}
\end{equation}
then the photon number noise in the
intervals $i$ and $j$
is nonclassically correlated.
Figures \ref{fig2}$(a)\!-\!(d)$ reveal
that the nonclassical correlation
of the $2$-soliton is substantially stronger than that of the
fundamental soliton even for an absorbing fiber.
Such an increase cannot be explained by a simple intensity scaling.
The effect is obviously related to the mentioned instability of
higher-order solitons. It is remarkable that
there exist propagation distances for which the nonclassical correlation
is stronger for an absorbing fiber than a nonabsorbing one.

The strongest photon number squeezing (smallest Fano factor)
achievable with an optimized broadband filter is illustrated
in Figs.~\ref{fig2}$(e)\!-\!(h)$.
Compared with the fundamental soliton, only a small increase
of the effect is observed for the $2$-soliton in the
$\omega$-domain
[Fig.~\ref{fig2}$(e)$, $6.6$ $\!\to$ $\!8.4$\mbox{dB}].
On the contrary, a rather strong increase of the effect can be observed
in the $x$-domain
[Fig.~\ref{fig2}$(g)$, $3.3$ $\!\to$ $\!9.6$\mbox{dB}],
provided that losses can be disregarded.
It is worth noting that the best photon number squeezing
is achieved in the $\omega$-domain for the fundamental soliton
and in $x$-domain for the $2$-soliton.
The results
show that the degree of squeezing
sensitively
depends
on the domain
considered.
Hence, replacing the Fourier transformation in Eq.~(\ref{eq.aw})
[including Eq.~(\ref{eq.filter})]
with more general transformation that relates the fields
in the two domains, may offer possibilities
of further optimization.
In particular, when we restrict our attention to linear
transformations which can be realized experimentally by
passive linear optical elements, then
we are left with a two-dimensional integral kernel function
to be optimized. In this way we may
hope that also for other nonlinear quantum objects
a considerable improvement of nonclassical features
can be achieved.

\vspace{2ex}
\noindent
{\bf Acknowledgment}\\
This work was supported by the Deutsche Forschungsgemeinschaft.

\vspace*{-2ex}




\vfill

\end{document}